\documentclass[journal]{IEEEtran}
\usepackage{amsmath,amsfonts}
\usepackage{algorithmic}
\usepackage{algorithm}
\usepackage{array}
\usepackage[caption=false,font=normalsize,labelfont=sf,textfont=sf]{subfig}
\usepackage{textcomp}
\usepackage{stfloats}
\usepackage{url}
\usepackage{verbatim}
\usepackage{graphicx}
\usepackage{cite}
\usepackage{xcolor}

\begin{document}

\title{Toward a Realistic Encoding Model of Auditory Affective Understanding in the Brain}

\author{Guandong Pan, 
        Yaqian Yang, 
        Shi Chen, 
        Xin Wang, 
        Longzhao Liu,
        Hongwei Zheng, 
        Shaoting Tang
        
    \thanks{Hongwei Zheng and Shaoting Tang are the corresponding authors.}
    
    \thanks{The authors are with School of Computer Science and Engineering, Beihang University, Beijing 100191, China, School of Artificial Intelligence, Beihang University, Beijing 100191, China, Hangzhou International Innovation Institute, Beihang University, Hangzhou 311115, China,
    Key laboratory of Mathematics, Informatics and Behavioral Semantics, Beihang University, Beijing 100191, China,
    Beijing Advanced Innovation Center for Future Blockchain and Privacy Computing, Beihang University, Beijing 100191, China,
    Zhongguancun Laboratory, Beijing 100094, China,
    State Key Laboratory of Complex \& Critical Software Environment, Beihang University, Beijing 100191, China,
    Institute of Trustworthy Artificial Intelligence, Zhejiang Normal University, Hangzhou 310012, China,
    Institute of Medical Artificial Intelligence, Binzhou Medical University, Yantai 264003, China,
    Beijing Academy of Blockchain and Edge Computing, Beijing 100085, China. Email:\{pan\_gd, yangyaqian, chenshi13579, wangxin\_1993, longzhao\}@buaa.edu.cn, hwzheng@pku.edu.cn,
     tangshaoting@buaa.edu.cn.}

}

\markboth{Journal of \LaTeX\ Class Files,~Vol.~14, No.~8, August~2021}%
{Shell \MakeLowercase{\textit{et al.}}: A Sample Article Using IEEEtran.cls for IEEE Journals}


\maketitle

\begin{abstract}
In affective neuroscience and emotion-aware AI, understanding how complex auditory stimuli drive emotion arousal dynamics remains unresolved. This study introduces a neurobiologically informed computational framework to model the brain’s encoding of naturalistic auditory inputs into dynamic behavioral/neural responses across three datasets (SEED, LIRIS, self-collected BAVE). Guided by neurobiological principles of parallel auditory hierarchy, we decompose audio into multilevel auditory features (through classical algorithms and wav2vec 2.0/Hubert) from the original and isolated human voice/background soundtrack elements, mapping them to emotion-related responses via cross-dataset analyses. 
Our analysis reveals that high-level semantic representations (derived from the final layer of wav2vec 2.0/Hubert)  exert a dominant role in emotion encoding, outperforming low-level acoustic features with significantly stronger mappings to behavioral annotations and dynamic neural synchrony across most brain regions ($p < 0.05$). Notably, middle layers of wav2vec 2.0/hubert (balancing acoustic-semantic information) surpass the final layers in emotion induction across datasets. Moreover, human voices and soundtracks show dataset-dependent emotion-evoking biases aligned with stimulus energy distribution (e.g., LIRIS favors soundtracks due to higher background energy), with neural analyses indicating voices dominate prefrontal/temporal activity while soundtracks excel in limbic regions.
By integrating affective computing and neuroscience, this work uncovers hierarchical mechanisms of auditory-emotion encoding, providing a foundation for adaptive emotion-aware systems and cross-disciplinary explorations of audio-affective interactions.
\end{abstract}

\begin{IEEEkeywords}
 Neural encoding, Naturalistic Auditory Stimuli, Neural Synchrony, Affective Understanding, Electroencephalogram (EEG).
\end{IEEEkeywords}

\section{Introduction}
Emotion serves as a critical window through which humans engage with the world. Quantifying how environmental stimuli drive emotion dynamics is vital for advancing affective neuroscience and emotion-aware AI \cite{damasio_descartes_2005,picard_affective_2000,meredithsomers_emotion_2019}. As a well-established approach, the naturalistic paradigm employs real-world audiovisual inputs to elicit time-varying emotional responses in the brain \cite{zheng_investigating_2015,jiang_multimodal_2023,baveye_lirisaccede_2015,koelstra_deap_2012}. Under naturalistic conditions, recent studies have made significant progress in recognizing/predicting emotions (e.g., discrete categories like positive or negative) from multimedia or physiological signals \cite{zheng_investigating_2015,jiang_multimodal_2023,baveye_lirisaccede_2015,koelstra_deap_2012}.
However, understanding how complex, naturalistic auditory stimuli dynamically modulate the fluctuations of emotion arousal is still a critical gap, which has been overlooked by traditional emotion recognition frameworks. 
Here, by integrating neurobiological principles of auditory processing with advanced computational modeling and multimodal data (stimuli data, behavioral annotations, EEG), we investigate how the brain encodes naturalistic auditory inputs into emotion arousal dynamics.

\begin{figure*}[htbp]
  \centering
  \includegraphics[width=\textwidth]{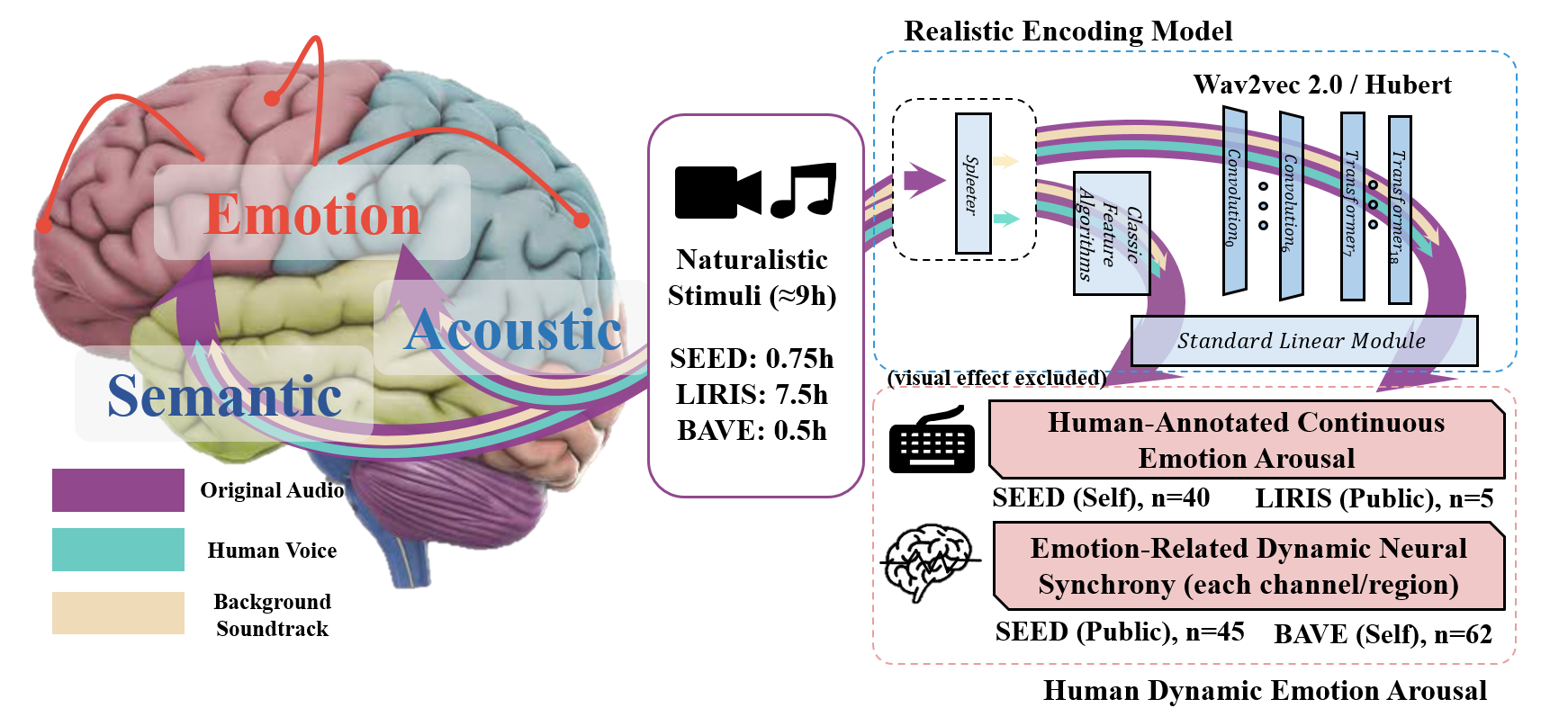}
  \vspace{-2mm}
  \caption{
  \textbf{Framework for quantitative modeling of auditory-driven emotion arousal encoding in the brain.}  We integrate naturalistic stimuli from datasets (SEED, LIRIS, BAVE; total \~9h) and analyze emotion-elicitation mechanisms of hierarchical/multi-element audio features (original audio, isolated human voice or background soundtrack) extracted via classical algorithms and deep networks (Wav2vec 2.0/Hubert). To isolate audio-specific effects, visual features (low-level attributes and high-level CLIP features \cite{radford_learning_2021}) are first incorporated as covariates during model training and excluded during prediction (following \cite{huth_natural_2016}). Using raw and source-localized EEG, we compute group-level emotion-related dynamic neural synchrony for each electrode/brain region. Finally, we map decomposed audio factors to human-annotated continuous emotion arousal or dynamic neural synchrony (evaluated via cross-validated Pearson correlation) to reveal how the brain encodes auditory information into emotion arousal dynamics.}
  \label{fig:framework}
  \vspace{-2mm}
\end{figure*}

The audio decomposition strategies employed in this work are grounded in the principles of hierarchical and parallel auditory processing \cite{li_dissecting_2023,hamilton_parallel_2021}, and attentional mechanisms \cite{rosenkranz_eegbased_2021,ferrante_adversarial_2025} in the human brain. Neuroscience research under the naturalistic paradigm has illuminated how the brain processes complex real-world inputs \cite{yang_default_2023}. For example, studies of speech processing \cite{li_dissecting_2023} reveal a multi-level auditory pathway—from subcortical structures like the auditory nerve to primary and non-primary auditory cortices—where lower-level regions (e.g., subcortical, primary cortex) handle basic acoustic features, while non-primary cortices specialize in abstract linguistic processing (semantics, grammar). Parallel processing of sound information across these hierarchical representations has also been demonstrated \cite{hamilton_parallel_2021}. Additional works on attentional focus and the global workspace theory \cite{rosenkranz_eegbased_2021,ferrante_adversarial_2025} highlight that in complex acoustic environments, neural activity correlates with selectively attended elements. Yet, how specific levels/elements of audio information contribute to emotion induction—both in terms of mechanism and magnitude—remains understudied. Building on classical  feature extraction algorithms\cite{wang_affective_2006,wang_video_2015,hu_eeg_2023} and deep learning-based methods \cite{baevski_wav2vec_2020,hennequin_spleeter_2020,radford_learning_2021}, we decompose audio into hierarchical (from acoustic to semantic) and multi-element (original audio, isolated voice, isolated soundtrack) components, then quantitatively model their associations with emotion-related responses while excluding the effect of visual features, as outlined in our research framework (Fig. \ref{fig:framework}).

Recent advanced deep neural network (DNN) models have provided powerful capabilities of semantic information representation and factor decomposition, and have been widely used in the encoding study for brain information processing \cite{oota_deep_2024}. Researchers \cite{schrimpf_neural_2021,goldstein_shared_2022,caucheteux_evidence_2023} have provided evidence that text-based language models and brain language processing both follow the predictive coding theory. On the basis of these studies, Li et al. \cite{li_dissecting_2023} and Millet et al. \cite{millet_realistic_2022} explored the commonalities between the processing of speech in the human brain and DNN models, and revealed that the hierarchy of DNN models is similar to the hierarchical speech processing mechanisms in the human brain. Therefore, it is of great significance to combine the powerful representational capacity of DNN models for the study of the encoding process of emotions. This work will utilize the advanced tool spleeter \cite{hennequin_spleeter_2020} to decouple audio stimuli into character voices and background soundtrack, and integrate the cutting-edge audio model wav2vec 2.0 \cite{baevski_wav2vec_2020} and Hubert \cite{hsu_hubert_2021} to represent the hierarchical audio information.

Second-scale dynamics of emotion arousal are a prerequisite to investigating how specific components within naturalistic stimuli  exert their influence on emotional responses. Moreover, under group-level analysis, neural activities or behavioral data can disregard individual variations \cite{khosla_cortical_2021},  facilitating our comprehension of the general mechanisms underlying the affective encoding of audio stimuli. In the field of neuroscience, neural synchrony has been demonstrated to be closely associated with collective emotional experience \cite{nummenmaa_emotions_2012,nummenmaa_emotional_2014}. The inter-subject correlation (ISC), initially introduced by Hasson et al. \cite{hasson_intersubject_2004}, has become a fundamental method for identifying neural synchrony among individuals. Compared with other imaging modalities such as functional magnetic resonance imaging (fMRI), electroencephalography (EEG) presents advantages owing to its high temporal resolution \cite{zheng_investigating_2015}, which is crucial for capturing subtle changes in neural responses. Multiple studies \cite{dmochowski_audience_2014,dmochowski_extracting_2018,kaneshiro_natural_2020} have verified that EEG-based neural synchrony is closely related to the audience preference or the structure of stimuli. Dmochowski et al.\cite{dmochowski_audience_2014,dmochowski_correlated_2012} explored dynamic neural synchrony and successfully linked it to movie scenes. Pan et al. \cite{pan_potential_2025} identified the reliability of single-electrode-based dynamic neural synchrony, allowing for a comparative analysis of regional differences in neural synchrony. Moreover, the raw EEG signals recorded from the scalp are a complex mixture of electrical activities originating from multiple neural sources within the brain. Source localization techniques can estimate the actual neural activities in different brain regions \cite{wang_identifying_2022}, which is beneficial for more accurately uncovering the neural mechanisms underlying emotion generation.
Therefore, in this paper, we extract dynamic neural synchrony from neural activities of different channels/regions (raw EEG electrodes or estimated brain regions) respectively to obtain emotion-related neural responses. The dynamic neural data of each channel/region and behavioral data at the group level are respectively modeled with audio features to study the general emotion arousal encoding process.

This study makes three key contributions by constructing a realistic encoding framework for the brain's auditory affective understanding. First, we demonstrate that semantic auditory representations (derived from the final layer of wav2vec 2.0/Hubert) exert a more dominant role in emotion encoding than low-level acoustic features, as evidenced by stronger mappings to behavioral annotations and dynamic neural synchrony across most brain regions. Second, middle layers of wav2vec 2.0 and hubert (balancing acoustic and semantic information) outperform the final layer in emotion induction, a finding robust across three datasets. Third, we show heterogeneous emotion-evoking effects of human voices and background soundtracks across datasets, with element dominance linked to stimulus-specific attributes (e.g., energy distribution).
These findings not only deepen our understanding of the underlying mechanisms but also pave the way for future research in this area, setting the stage for further exploration and application of audio-emotion interactions.

\section{Materials and Methods}\label{sec:methods}
\subsection{Datasets} \label{sec:dataset}
\begin{figure} 
    \centering
    \includegraphics[width=0.8\linewidth]{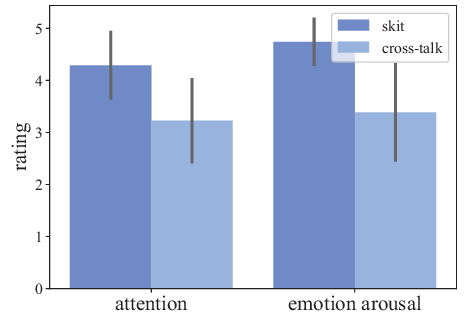}
    \caption{\textbf{Subject ratings for the used stimuli in BAVE dataset.} The scoring is on a five-point scale. Each subject scored immediately after watching the corresponding short video. Error bar represents the 95\% confidence interval of ratings.}
    \label{fig:rating} 
\end{figure}

This work utilizes two emotion-related public datasets, one self-collected annotation dataset, and one self-collected EEG dataset. Public datasets: (i) the SEED EEG dataset \cite{zheng_investigating_2015}, containing 62-channel EEG data from 45 experiments with 15 approximately 4-minute video clips, and (ii) the LIRIS video dataset \cite{li_continuous_2015} with 30 full movies (10 minutes–1 hour) annotated for arousal by 5 annotators. For emotion arousal annotation of SEED video clips, we recruited 1061 individuals to participate in the experiment. This resulted in 600 valid annotation sequences, with each movie clip getting 40 continuous annotations. Additionally, the EEG dataset (Beihang Audio-Visual EEG dataset, BAVE) was self-collected from 65 subjects (3 excluded due to data quality issues) while they viewed three comedy videos: a cross-talk (audio-only), a mime (visual-only) and a skit (audio-visual), each approximately 15 minutes in duration. Each participant in the experiment received a reward of 150 yuan. Only the cross-talk and skit are utilized in this study. The used sample comprised 43 males and 19 females, with ages ranging from 18 to 34 years (M = 22.16, SD = 9.15). EEG signals were recorded at a 1000 Hz sampling rate using a 64-channel ESI Neuroscan System (with M1 and M2 as reference channels, resulting 62-channel EEG data), along with resting-state sessions and post-video self-assessment questionnaires. Fig. \ref{fig:rating} shows the attention level and emotion arousal level of the audience's ratings for the two videos. The average scores higher than 3 indicate that the two videos have high attractiveness. All procedures were ethically approved. Detailed collection procedure see Supplementary Appendix I.A.

\subsection{EEG data preprocessing}
The preprocessed EEG data collection of SEED is used in our work, which has been sampled at 200 Hz and filtered from 0-75 Hz. We have performed the same operation on our self-collected BAVE EEG dataset. Moreover, we conduct further preprocessing using EEGLAB toolbox \cite{delorme_eeglab_2004} in MATLAB. A notch filter (48-52 Hz) is applied to remove power line interference. 
Eye movements and muscle artifacts are decomposed and removed by the independent component analysis (ICA) technique in EEGLAB. 

\textbf{Source localization.} Brain source localization maps scalp-recorded EEG signals to underlying cortical activity, enabling emotion-related neural correlates to be localized in brain regions. We perform source localization using the MNE library \cite{gramfort_meg_2013}, following a standardized pipeline: 1) constructing the source space and boundary element model (BEM) for brain tissue conductivity; 2) configuring electrode montages and reference channels for each EEG dataset; 3) computing forward solutions to link brain sources with scalp EEG signals, estimating noise covariance matrices, generating inverse operators, and applying dynamic statistical parametric mapping (dSPM) to raw data for source activity estimation. Anatomical time-series are extracted using the aparc parcellation scheme \cite{dale_cortical_1999}, yielding 68 left/right hemisphere regions. For visualizations of electrode positioning and MRI alignment across datasets, see Supplementary Appendix I.F.

\subsection{Emotion-related response acquisition} \label{sec:Emotionacquisition}

\textbf{Dynamic neural synchrony extraction.} We utilize the state-of-the-art dynamic neural synchrony method in EEG \cite{pan_potential_2025}, which measures the similarity of single-channel brain activities among  the subjects in the present study (exposed to identical naturalistic stimuli). This approach has the advantage of capturing shared cognitive information without predetermined hypotheses and isolating heterogeneous cognitive processes from different channels/cerebral cortices for regional comparison. Previous studies have shown that neural synchrony is closely related to emotion \cite{nummenmaa_emotions_2012,dmochowski_audience_2014}, hence we use dynamic neural synchrony here to reflect changes related to emotion arousal. 

We apply this technique to both EEG datasets, each including both the raw EEG data ($eeg$) and the time-series of source activity ($brain$). 
The implementation method is briefly introduced below; for details, refer to \cite{pan_potential_2025}. We measure the similarity of each pairwise time sequences under each window using a sliding window method for each video clip and each channel. By averaging the dynamic correlations of all pairwise subjects under the corresponding windows, we obtain a population-level dynamic neural synchrony sequence for each clip. Pearson correlation coefficient (PCC) is used for similarity measurement, and the window size is set to 10 seconds (see Supplementary Appendix II.A for sensitivity analysis of the sliding window size), and the step size is 1 second. Before correlation, EEG is processed as the first-order differential feature to improve performance, following \cite{pan_potential_2025}. The Python library \emph{taichi} \cite{hu_taichi_2019} is used to compute sequences in parallel to accelerate the computation.

\textbf{Human annotated dynamic emotion arousal.} LIRIS dataset provides a 1 Hz continuous emotion arousal label for each movie clip by 5 participants. For the SEED dataset, we used the online platform Naodao to publicly recruit a group of Chinese native people to watch movie materials in SEED twice and collect their annotation in the second time, namely the post-hoc manner \cite{ding_interbrain_2021}. The annotators reported their dynamic emotion arousal using a slider in the website. The experiment program was designed by Psychopy \cite{peirce_psychopy2_2019}. Each movie clip was reported by 40 different annotators. We paid each annotator 3 yuan for each 4-min around film clip. An annotation quality check module was used to ensure data quality. We average annotation sequences of each clip across annotators to gain the final continuous emotional responses. Detailed collection procedure and data processing are available in Supplementary Appendix I.A.

\begin{figure}[htbp] 
    \centering
    \includegraphics[width=\linewidth]{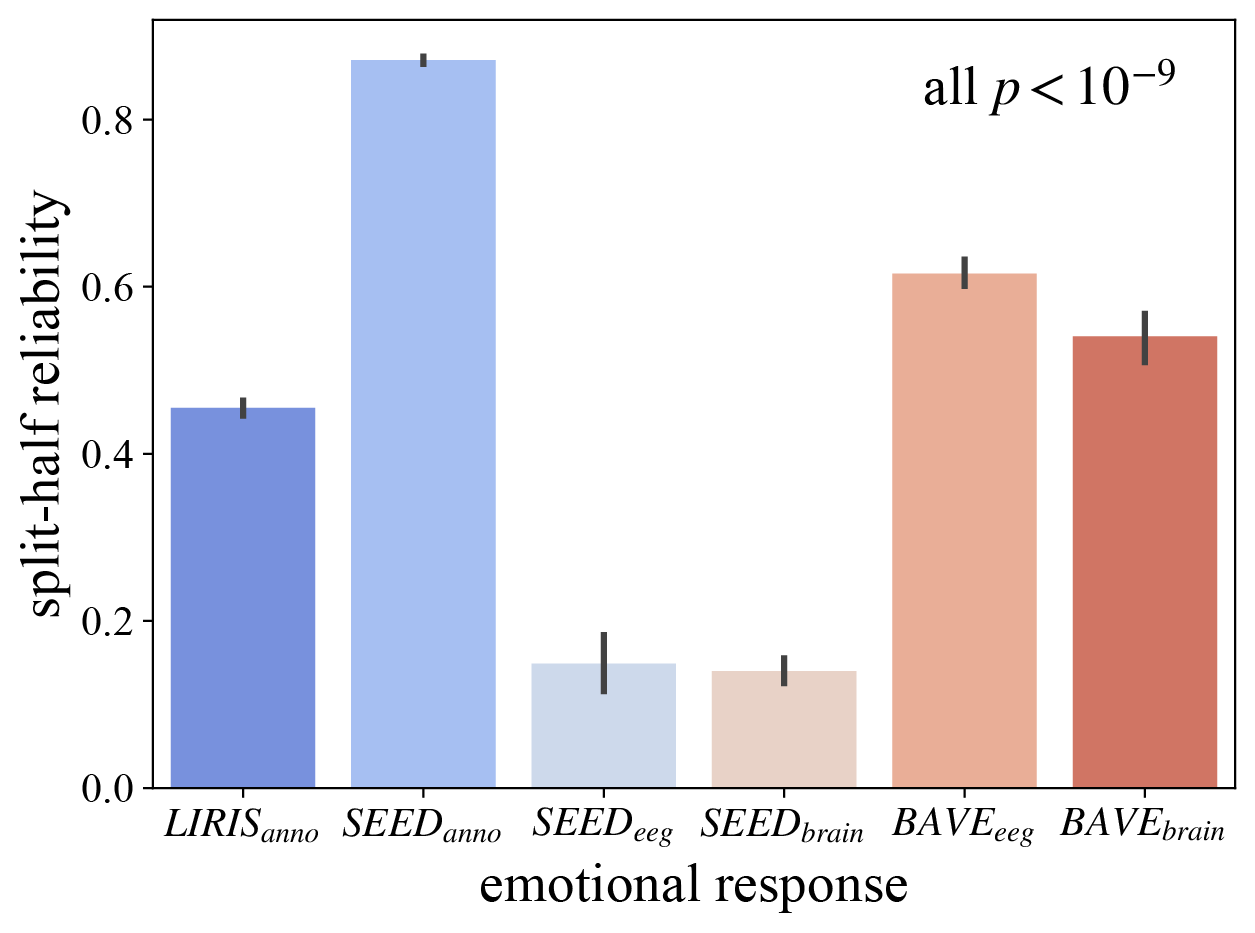}
    \caption{\textbf{Split-half similarity of emotion-related responses across datasets (SEED, LIRIS, BAVE) and modalities (behavioral annotations, raw EEG, source-localized brain activity).} To assess the reliability of behavioral annotations and dynamic neural synchrony through raw EEG signals or source-localized brain activity, we perform 100 rounds of randomly splitting subjects into two groups and compute the Pearson correlation coefficient (PCC) of continuous emotion-related responses between groups. All comparisons exhibit statistically significant non-zero correlations ($p<10^{-9}$). Error bar represents the 95\% confidence interval of the averaged similarity.}
    \label{fig:reliability} 
\end{figure}

\textbf{Split-half similarity of emotion-related responses.}
To validate group-level reliability of emotion-related responses, we perform 100 rounds of randomly splitting the subjects into two separate groups, comparing intergroup similarity of behavioral/neural dynamics. All measures—human annotations, raw EEG, and brain source activity—show non-zero correlations ($p<10^{-9}$, Fig. \ref{fig:reliability}), confirming emotion-related response consistency across subjects.

\subsection{Characteristics of naturalistic stimuli} \label{sec:extraction_audio}
\textbf{Features from classical feature extraction algorithms.} Acoustic features reflect core sound properties (e.g., pitch, timbre), and their capacity to discriminate emotional information has been widely acknowledged in prior speech emotion recognition research \cite{wang_video_2015}. In this work, we extract 12 low-level descriptors (LLDs) to characterize acoustic information: \emph{logarithm energy spectrum (LES)}, \emph{short-time energy (STE)}, \emph{zero-crossing rate (ZCR)}, \emph{sound pressure level (SPL)}, along with their first-order difference features and second-order difference features. These LLDs are computed for each one-second frame. Detailed mathematical formulations of the LLDs are provided in Supplementary Appendix I.B. The considerations for feature selection are also provided in Supplementary Appendix I.C.

\textbf{Representations derived from deep learning.} Previous research has shown that activations of DNNs can deliver rich semantic information compared to traditional feature extraction methods \cite{caucheteux_evidence_2023,huth_natural_2016,caucheteux_brains_2022}. Recent studies  \cite{millet_realistic_2022,vaidya_selfsupervised_2022} based on the self-supervised speech model (e.g., wav2vec 2.0 \cite{baevski_wav2vec_2020} and Hubert \cite{hsu_hubert_2021}) have established links between model activations and brain dynamics. Following this paradigm, we employ both wav2vec 2.0 and Hubert to map audio representations to emotions, using pre-trained weights from \cite{zhang_wenetspeech_2022}.
Both models share a similar architecture: wav2vec 2.0 consists of a 7-layer convolutional feature encoder and 12 transformer blocks \emph{f} : $\mathcal{X} \mapsto \mathcal{C}$, taking raw audio $\mathcal{X}$ (16 kHz waveform) as input and outputting semantic representations $c_1,c_2,...,c_T$ \cite{baevski_wav2vec_2020}; Hubert follows a comparable design but with enhanced contrastive learning for speech units \cite{hsu_hubert_2021}. The transformer layer output  $\mathcal{C}$ has a 768-dimensional feature space (50 Hz frequency, 20 ms stride), while the convolutional layer yields 512-dimensional features. To ensure fair comparison with low-level descriptors (LLDs), we apply PCA to reduce semantic features to 12 dimensions (see Supplementary Appendix II.B for sensitivity analysis of dimensionality reduction). Supplementary Appendix I.D describes the characterization of wav2vec 2.0 representations in detail. Briefly, semantic features tend to derive primarily from the 12-layer Transformer blocks, while their 7-layer convolutional encoders appear to capture low-level acoustic representations.

\textbf{Element separation.} Guided by the neurobiological principle of parallel processing \cite{hamilton_parallel_2021}, we decompose audio into human voice and background elements, as the brain’s auditory cortex processes speech and non-speech sounds through distinct neural pathways. To isolate human voices and background soundtracks from audio clips across the SEED, LIRIS, and BAVE datasets, we employ the pre-trained 2-stem model of the Python library \emph{Spleeter} \cite{hennequin_spleeter_2020}—a state-of-the-art tool specialized for vocal-background audio separation. Specifically, human voice components encompass character dialogues, monologues, and voice-overs, while background soundtracks comprise accompanying music and sound effects.
The robustness of this separation is validated via qualitative audio demonstrations, human perceptual evaluations (e.g., Likert-scale clarity ratings), and cross-model comparisons (vs. alternative separators like Open-Unmix \cite{stoter_openunmix_2019}) on the SEED dataset, with detailed results provided in Supplementary Appendix I.E. Multilevel features (as introduced in Sections on classical algorithm-derived features and deep learning-derived representations) are further extracted from the two separated audio elements for subsequent emotion encoding analyses.

\textbf{Visual features.} To better isolate audio-specific effects and enable an unbiased analysis, we extract both classical low-level visual features and dimensionality-reduced high-level visual features, incorporating them as covariates for model estimation to account for their contribution to emotion-related responses. These visual covariates are subsequently excluded during the prediction phase to emphasize the independent influence of auditory features. This methodological approach is consistent with the strategy utilized by Huth et al. \cite{huth_natural_2016}.
For low-level visual features, we compute classical attributes on a per-second basis, including color, brightness, and texture features, among others \cite{baveye_affective_2018}; detailed extraction procedures are provided in Supplementary Appendix I.B.

For high-level visual features, we utilize the CLIP model (openai/clip-vit-base-patch32) \cite{radford_learning_2021}, a vision-language model proven effective in capturing abstract visual semantics. The 768-dimensional high-level visual features extracted by this model are reduced to 12 dimensions via PCA, matching the dimensionality of our high-level audio features (such as $H_{PCA}$). These reduced high-level visual features are concatenated with the low-level visual features to form a combined covariate set, which is included during model training. In the prediction phase, however, our focus remains exclusively on the contribution of audio features.

\subsection{Emotion score} \label{sec:emotionscore}
After extracting auditory features, visual features and emotion-related responses, we explore how audio factors are encoded into emotion arousal dynamics. The emotion score, calculated through systematic methodological steps, serves as a pivotal indicator in this context: it directly quantifies the performance of audio information encoding by measuring the correlation between predicted emotion-related responses (derived from model prediction) and actual emotion-related responses  (i.e., behavioral annotations or raw/source-localized dynamic neural synchrony). This metric is motivated by the need to rigorously assess how well auditory features capture and translate into emotional content, with its validity underpinned by a series of methodological procedures and related encoding literature \cite{caucheteux_evidence_2023,millet_realistic_2022,huth_natural_2016}.

\textbf{Temporal alignment.} Dynamic neural synchrony uses a 10s sliding window, which results in its sequence length being smaller than the video length. Therefore, to ensure the same sample length, we perform the sliding window process on all other kinds of data and average the data of each window. Step size is set to 1 second. Among them, high-level features are aligned by intercepting the input audio with n sliding windows and averaging the output representation of each window across the time scale.

\textbf{Data partition.} In this work, we mainly use the stratified five-fold cross-validation method to evaluate the performance of the encoding. Specifically, in each evaluation, the stratified partition randomly divides one-fifth of multi-modal samples of each video into one of the five subsets. Each subset is used as a test set once, while the other four subsets are used as the training set. The average score of all 5 subsets as test set can be regarded as one emotion score. 

\textbf{Penalized linear regression.} Following \cite{caucheteux_evidence_2023,millet_realistic_2022,huth_natural_2016}, we build the linear mapping with $l_2$-penalized standard linear model (Ridge Regression) to encode emotion-related responses given the audio features and visual features (see Supplementary Appendix II.D for linear-nonlinear comparison). These emotion-related responses include behavioral data, dynamic neural synchrony for each electrode, and dynamic neural synchrony for each brain region after source localization, with each computed separately. Data standardization is applied to the samples. The RidgeCV function \cite{pedregosa_scikitlearn_2011} is used with the penalization parameter $\lambda$ ranging from $10$ to $10^8$ (20 values scaled logarithmically). $\lambda$ is independently chosen for each model with a \textbf{nested} five-fold cross-validation over the train set. The formula of the optimization is the following:
\begin{equation}
{\arg\min}_{V} \sum_{i \in train_s} \left( V^\top \hat{X}_i-y_i \right)^2 + \lambda \left\lVert V \right\rVert^{2} \quad 
\end{equation}
where $V$ denotes the weight vector of the linear model, $\hat{X}_i$ represents the concatenated feature vector (auditory + visual features) of the $i$-th sample, and $y_i$ is the corresponding emotion-related response (behavioral annotation or dynamic neural synchrony).

\textbf{Evaluation.} Assuming the linear mapping as $W$, we quantify the correlation between $W \cdot X_{test}$ and $Y_{test}$  to evaluate the encoding ability from auditory features to emotion-related responses. Specifically, we extract visual features and incorporate them as covariates in our encoding models: during training, visual features are concatenated with auditory features to account for their potential influence; during prediction, visual feature values are set to 0 to isolate the unique contribution of auditory features. This approach, aligned with Huth et al. \cite{huth_natural_2016}, directly quantifies audio-specific effects by controlling visual confounds. To further validate the independence of auditory contributions and characterize visual modality's impact on emotion encoding, we designed ablation experiments across four modality combinations (detailed in Supplementary Appendix II.E), comparing auditory-only, visual-only, and covariate-integrated models.

\begin{equation}
    Score = PCC (Y_{test}, W \cdot X_{test})
\end{equation}

where PCC means Pearson correlation coefficient.

\subsection{Stepwise regression}\label{sec:stepwise}
In the joint modeling of acoustic and semantic features from the original audio, along with semantic features from character voices and soundtracks, we employ stepwise regression to systematically quantify the incremental contributions of these features to encoding capabilities of emotion arousal. Given the high dimensionality of semantic features, we first apply PCA to the semantic features (to reduce dimensionality to 12 dimensions, see Supplementary Appendix II.B for sensitivity analysis of dimensionality reduction) prior to conducting the analyses.

Specifically, we concatenate the two corresponding feature groups (acoustic and semantic, or voices and soundtracks) and train the model iteratively by incrementally adding one feature at a time, starting from a single feature. Each training iteration utilizes five-fold cross-validation to compute emotion scores. To mitigate the risk of spurious significance arising from arbitrary variable ordering, a known concern with stepwise regression \cite{mundry_stepwise_2009}, we randomize the order of variables within each feature group and repeat the entire stepwise analysis 100 times. This procedure aligns with similar practices in neuroscience for dissecting incremental feature effects \cite{yang_enhanced_2023}.

Furthermore, manipulating the order in which the two feature groups are added to the encoding model allows for explicit quantification of the incremental contributions of hierarchical features, as well as comparisons between voice and soundtrack features.

\subsection{Statistics} \label{sec:statistics}
In split-half experiments examining emotion-related responses, we employ the Wilcoxon signed-rank test to determine if the distribution of split-half similarity (e.g., inter-group correlation coefficients) deviates significantly from zero. A non-significant result would suggest that the observed consistency in emotion-related responses could be ascribed to random chance. Conversely, a significant outcome corroborates that the similarity across groups reflects non-random reliability inherent to emotion-related processing.

For significant-mapping experiments, the Mann-Whitney U test is utilized to identify annotations/EEG channels/source-localized brain regions where the empirical distribution (real data) differs significantly from the null distribution.
Regarding neural mappings, to account for the increased likelihood of Type I errors due to multiple comparisons, we apply the Benjamini/Hochberg False Discovery Rate (FDR) correction \cite{benjamini_discovering_2010}. This approach balances the need to detect true effects while controlling the proportion of falsely rejected null hypotheses, making it well-suited for exploratory analyses where identifying potential effects is prioritized.
In experiments comparing score differences across brain regions, the Mann-Whitney U test is used to assess between-group differences in emotion scores for each region. Here, we employ the Family-Wise Error Rate (FWER) correction to control the probability of making at least one Type I error across all tested regions.

\begin{figure*}[htbp] 
  \centering
  \includegraphics[width=\textwidth]{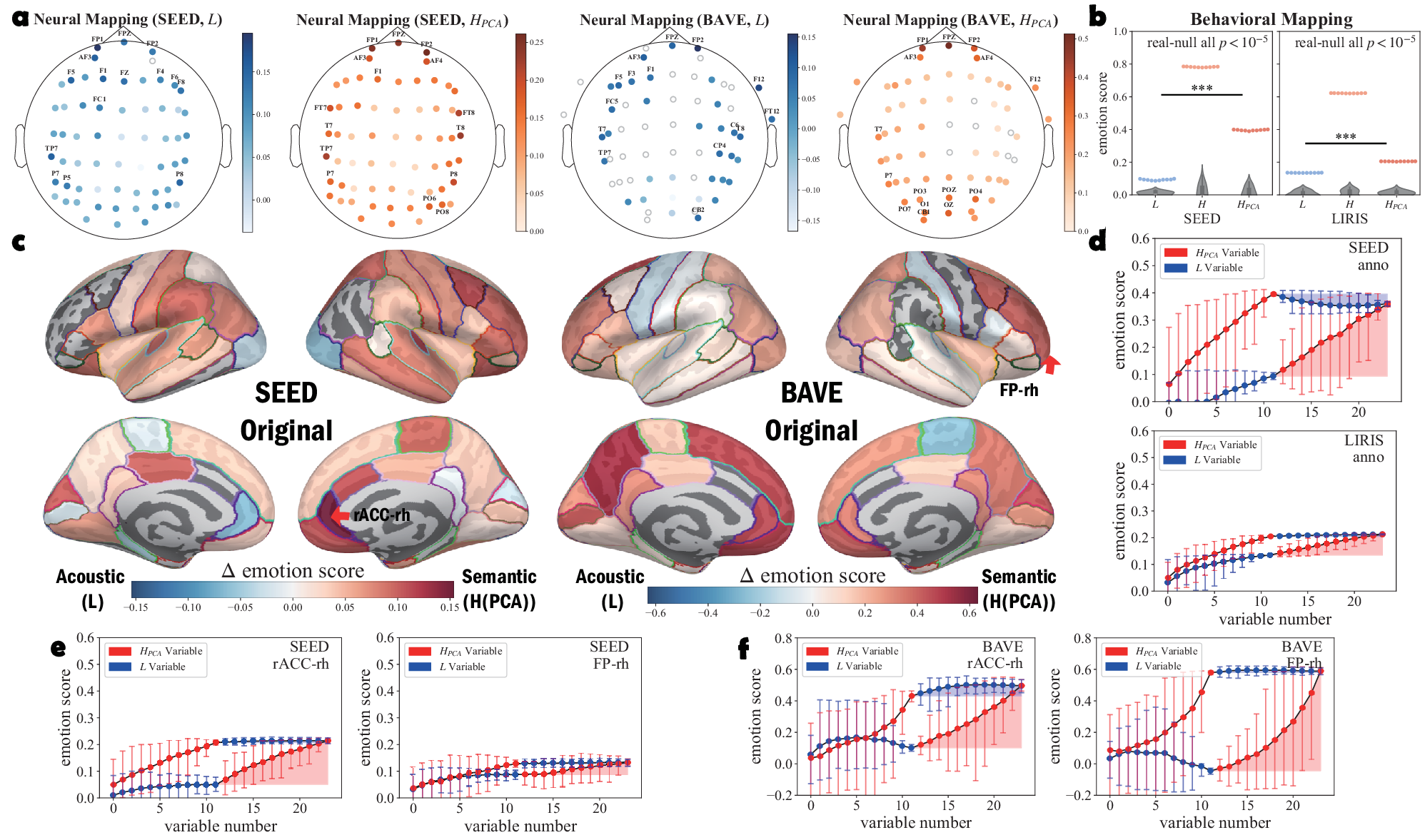}
  \caption{\textbf{The dominant role of semantic representations compared to acoustic features.} \textbf{a} and \textbf{b} illustrate the significant mappings from classical algorithms features ($L$, low-level acoustic features) and the wav2vec 2.0 final layer representations ($H$, high-level semantic features; or PCA-reduced $H_{PCA}$) to responses of behavioral annotations($p<10^{-5}$) or dynamic neural synchrony of raw EEG signals ($p<10^{-3}$). Significance is computed between real distribution and null distribution by shuffling responses (neural data corrected by FDR). Channels that fail the test in \textbf{a} are marked as a hollow circle. In \textbf{b}, emotion scores of semantic features (both $H$ and $H_{PCA}$) is significantly higher than acoustic features ($p<0.001$). \textbf{c} shows the score difference between hierarchical audio features of SEED and BAVE datasets. The scores are computed between audio features and dynamic neural synchrony of brain source activities. Only significant differences are colored ($p<0.05$, corrected by FWER). \textbf{d-f} Stepwise regression of joint low-level acoustic ($ L $) and semantic ($ H_{PCA} $) features predicting emotion scores (behavioral: \textbf{d}; brain regions in SEED: \textbf{e}; brain regions in BAVE: \textbf{f}). Whiskers reflect variability (box-plot conventions).  Across 100 iterations of randomized feature addition, red shading shows emotion scores improved when $ H_{PCA} $ was added after $ L $, while blue shading shows minimal change when $ L $ was added after $ H_{PCA} $, demonstrating semantic features’ incremental value for emotion encoding.}
  \label{fig:high_low}
\end{figure*}

\section{Results} \label{sec:results}

\subsection{Multilevel auditory features of the original audio and the isolated elements can be significantly mapped to emotion-related responses}

We conducted cross-dataset experiments to explore whether multilevel auditory features associate with behavioral and neural emotion-related responses. For dynamic neural synchrony, Fig. \ref{fig:high_low}a and Fig. \ref{fig:bgm_vocal}a visualize emotion scores quantifying the link between auditory features (low-level descriptors, $L$, the reduced last layer representations of the wav2vec 2.0 model, $H_{PCA}$) and dynamic neural synchrony of raw EEG signals across 62 electrodes. Electrodes with empirical scores exceeding null distributions (model training with shuffled labels, $p<10^{-3}$, FDR-corrected) are denoted by solid circles; others appear as hollow circles. The top 15 electrodes (by score) in each subplot are labeled, highlighting significant activity in temporal, prefrontal, and occipital regions. Source localization analyses in Supplementary Fig. 5 and Fig. 6 further show colored brain regions with emotion scores differing significantly from null distributions, demonstrating widespread neural engagement with auditory features.

For behavioral annotations, Fig. \ref{fig:high_low}b presents the results of modeling three audio feature types ($L$, $H$ and $H_{PCA}$) from original audios alongside continuous arousal ratings or shuffled ratings in SEED/LIRIS, with emotion scores derived through this modeling process. In contrast, Fig. \ref{fig:bgm_vocal}b compares the emotion scores of voice/background elements against their respective null distributions to assess statistical significance. All feature types exhibit significant encoding of emotional signals, with behavioral scores surpassing null expectations ($p<10^{-5}$ for both datasets and three audio elements).

In addition to the final layer of wav2vec 2.0 model ($H$ and $H_{PCA}$), we also investigated the emotion-related associations of audio features from each layer of wav2vec 2.0 (full-dimensional, unreduced) with behavioral and neural responses. Behavioral results (Fig. \ref{fig:layerwise}a) and source-localized dynamic neural synchrony (Fig. \ref{fig:layerwise}c, \ref{fig:layerwise}d) depict emotion scores across model layers in the original audio, with null distributions visualized as gray bar plots/violin plots. Layerwise wav2vec 2.0 features extracted from human voice and background music elements are presented in Supplementary Fig. 7.
Results show significant emotion scores compared with null distribution across all layers in three audio elements, confirming hierarchical emotional representation where both shallow acoustic and deep semantic features contribute to emotion-related neural/behavioral responses.

\subsection{The dominant role of semantic representations over acoustic features in emotion encoding across most brain regions}

Considering the hierarchical and parallel processing mechanisms of auditory perception \cite{li_dissecting_2023,hamilton_parallel_2021}, we intend to investigate which level of auditory information has a stronger effect on inducing emotions. Here, we first compared the manually extracted low-level features ($L$) with the semantic features of the last layer of the wav2vec 2.0 model ($H$, $H_{PCA}$). Supplementary Fig. 4 shows that reducing wav2vec 2.0 features from 768 to 12 dimensions preserves critical variance information.
Fig. \ref{fig:high_low}b shows that both full-dimension and PCA-reduced semantic features ($H$, $H_{PCA}$) exhibit significantly higher emotion scores than acoustic features ($L$) on the behavioral mapping, with $p < 0.001$ for both SEED and LIRIS datasets. Additionally, across SEED and BAVE datasets, we calculated score differences between semantic and acoustic features for each source-localized brain region, statistically assessing significance ($p < 0.05$, FWER-corrected). Fig. \ref{fig:high_low}c highlights regions with significant differences: in most cortical areas, semantic features outperform acoustic features. The right hemisphere's rostral anterior cingulate cortex (rACC-rh, $\Delta score=0.15$) and frontal pole (FP-rh, $\Delta score=0.62$) show the highest score differences in SEED and BAVE respectively,  prompting their selection for subsequent analysis.

Then, we used stepwise regression (detailed in Section \ref{sec:stepwise}) \cite{yang_enhanced_2023} to explicitly quantify the incremental contribution of hierarchical features on emotional encoding. Briefly, we concatenated two feature groups ($L$ and $H_{PCA}$) and iteratively added features to compute emotion scores. The order of variables within each feature group were randomized and repeated 100 times. Fig. \ref{fig:high_low}d-f depict the results between joint hierarchical features and emotion-related responses across three datasets. Results show that when acoustic features are added later, emotion scores plateau rapidly in both behavioral and neural mapping analyses (small blue shaded areas), whereas adding semantic features later lead to significant improvements (red shaded areas). This pattern indicates that semantic features drive emotion encoding beyond acoustic cues.

\begin{figure*}[ht] 
  \centering
  \includegraphics[width=\textwidth]{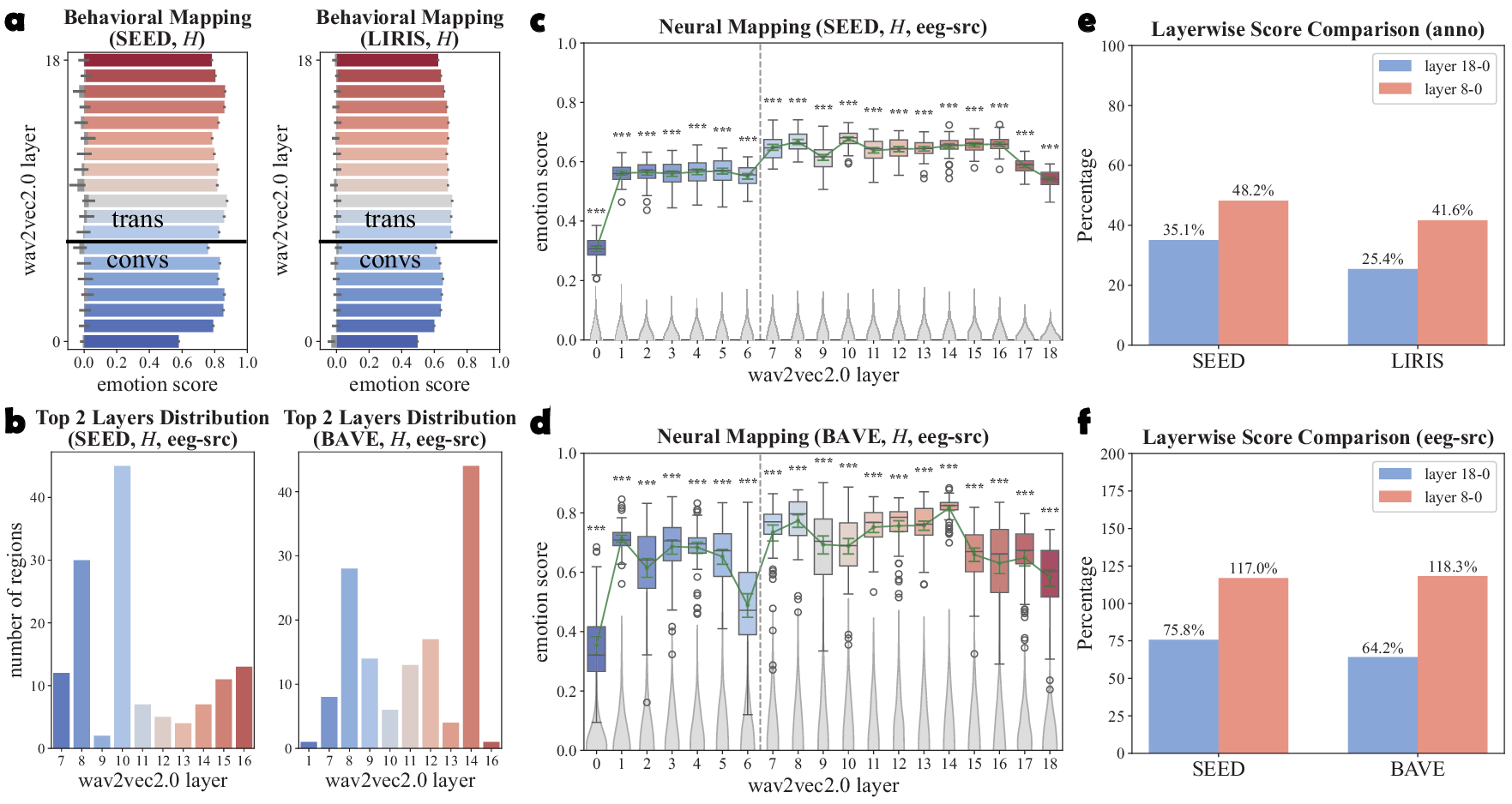}
  
  \caption{\textbf{Layerwise exploration of wav2vec 2.0 features (full dimensions) from the original audio in emotional encoding, integrating behavioral, neural, and cross-layer comparative results.}  \textbf{a} displays empirical and null emotion scores of behavioral mapping (SEED, LIRIS) across 19 layers (all $p<10^{-5}$): shallow/acoustic-dominated layers have low scores, middle layers (semantic-acoustic integration) show a sharp rise, and deep/semantic-dominated layers slowly descend. \textbf{b} Top 2 layers distribution (BAVE, SEED) depicts frequency of layers ranking top 2 in 68 brain regions, with middle layers (e.g., 8, 10, 14) dominant, reflecting neural preference for multi-scale integration. \textbf{c,d} Neural mapping (SEED, BAVE) shows emotion scores across layers (gray violins as null distributions, all $p<0.001$), with non-linear trajectories matching behavioral results. \textbf{e,f} Layerwise score comparison shows middle layers (e.g., layer 8) outperform deep layers (e.g., layer 8 scores 48.2\% higher than layer 0 vs. layer 18 scores 35.1\% higher than layer 0 in SEED), highlighting them as a “synergistic zone” for emotion encoding.}
  \label{fig:layerwise}
\end{figure*}

Additionally, we extended our analysis to compare emotion scores between the full-dimensional final layer and first layer of the wav2vec 2.0 model. Consistent with prior studies showing that wav2vec 2.0's shallow layers prioritize acoustic features while deep layers encode semantic information \cite{pasad_layerwise_2022,shah_what_2021}, we hypothesized that deep-layer representations would outperform shallow-layer features in emotional encoding. Comparative results in Fig. \ref{fig:layerwise}e and Fig. \ref{fig:layerwise}f demonstrate that average emotion scores from semantic representations (final layer) across brain regions are significantly higher than those from acoustic representations (first layer) at least 25\% across all datasets.

Collectively, these results demonstrate the dominant role of semantic features in emotional encoding than low-level acoustic features.

\subsection{Middle layers of wav2vec 2.0 and hubert demonstrate enhanced emotion induction compared to the final layer}\label{sec:middle_layer}

Having confirmed that high-level semantic features (the final layer of wav2vec 2.0) outperform acoustic features (either handcrafted or from the model’s first layer) in emotion encoding, we aim to probe deeper into the hierarchical representation space with the full feature dimension. This exploration is motivated by the nature of self-supervised models like wav2vec 2.0, which progressively transition from encoding primarily acoustic to semantic information across layers \cite{pasad_layerwise_2022,shah_what_2021,vaidya_selfsupervised_2022}. Given this hierarchical trajectory, the question of which specific auditory information optimally supports emotion encoding remains unresolved.

Layer-wise analysis of wav2vec 2.0 across all 19 layers (Fig. \ref{fig:layerwise}a, \ref{fig:layerwise}c, \ref{fig:layerwise}d) shows that emotion scores follow a non-linear path (see Supplementary Fig. 7 for the isolated elements, see Supplementary Fig. 12 for hierarchical effects of categorized lobes). They are low in shallow, acoustic-dominated layers, rise sharply in middle layers (e.g., layers 7–14), and plateau in deep, semantic-dominated ones. To validate the generalizability of this layer-wise trend across self-supervised audio models, layer-wise analysis of Hubert (Supplementary Fig. 10) also exhibits a consistent trend, with emotion scores showing analogous non-linear dynamics across its layers. Further analysis in Supplementary Fig. 11 controls for the full visual feature space (not just low-level visual features and high-level dimension-reduced subsets), confirming consistent auditory layer trends. 
Looking at the top 2 layers in 68 source-localized brain regions (Fig. \ref{fig:layerwise}b), middle layers (especially 8, 10 and 14) are dominant. Layer 8, 10, and 14 rank top in over 25 regions, showing neural preference for these layers in integrating multi-scale auditory information (see Supplementary Fig. 13 for similar results of top 1-5 layers distribution). 
Moreover, layer-wise score comparisons (Fig. \ref{fig:layerwise}e, \ref{fig:layerwise}f), normalizing to layer 0 (layer 0 refers to the first layer of the convolutional feature encoder, following \cite{millet_realistic_2022}), reveal middle layers (e.g., layer 8) outdo the final layer (layer 18). In SEED behavioral mapping, layer 8 scores are 48.2\% higher than layer 0, beating layer 18’s 35.1\%, a pattern consistent across datasets. Notably, the middle layers (7–14) of wav2vec 2.0/Hubert maintain the same feature dimensions to the final layer (768D), ruling out dimensionality as a confounding factor for their superior emotion induction.

Briefly, wav2vec 2.0’s middle layers (7–14) balance acoustic detail and semantic abstraction, driving stronger emotion encoding in behavioral and neural responses, even surpassing the deepest semantic layers. This refines our understanding: self-supervised models encode emotions most powerfully in intermediate, integrative representations, a key insight for future speech-emotion and neurocomputational work.

\subsection{The representations of human voices and background soundtracks exhibit heterogeneous emotion-evoking effects on three datasets}


When the human brain processes complex naturalistic auditory stimuli, multiple elements may contribute to emotional induction; however, the relative dominance of specific elements in driving emotional responses remains unclear. To address this, we first employed the deep neural network model \emph{Spleeter} \cite{hennequin_spleeter_2020} to isolate two primary components: human voices and background soundtracks. We then extracted audio representations from these isolated elements and modeled their relationships with emotion-related responses to compare their differential capacities for emotion induction.

Based on our prior finding that layer 8 of wav2vec 2.0 optimally balances acoustic and semantic information (Section \ref{sec:middle_layer}), we focus on layer 8 features to compare the emotion-evoking effects of voices and soundtracks (full dimension $H8$ or with PCA reduction $H8_{PCA}$). Comparative results for the model's final layer across these elements are provided in Supplementary Fig. 14.

\begin{figure*}[ht] 
  \centering
  \includegraphics[width=\textwidth]{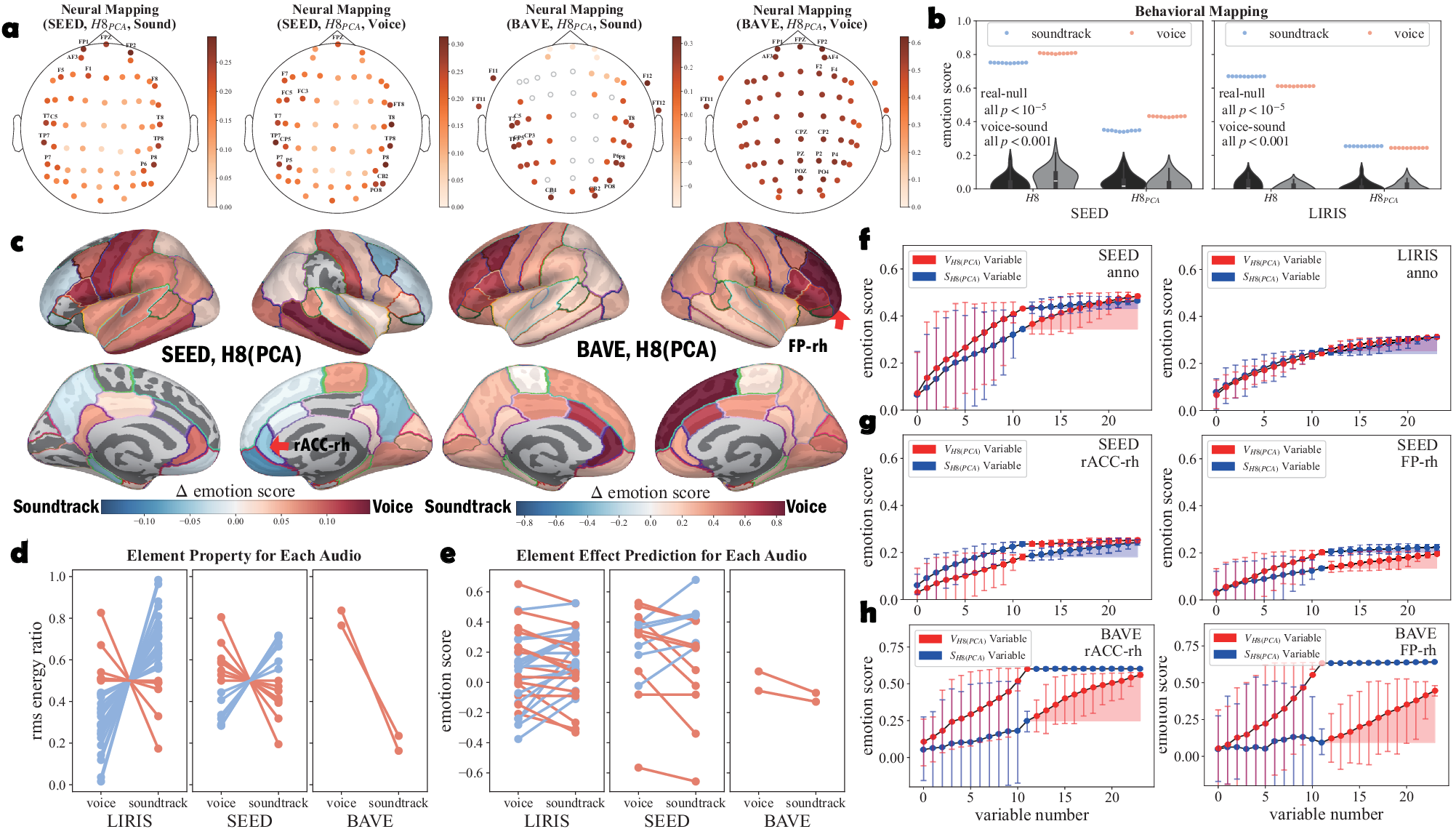}
  \caption{\textbf{Heterogeneous induction effect of the human voice and soundtracks.} \textbf{a} and \textbf{b} show the significant mappings of two isolated elements between the wav2vec 2.0 layer 8 representations ($H8_{PCA}$) and dynamic neural synchrony of raw EEG signals ($p<10^{-3}$) or behavioral annotations ($p<10^{-5}$) on three datasets (see Supplementary Fig. 14 for results of wav2vec 2.0 final layer). Significance computation and illustration is similar to Fig. \ref{fig:high_low}a–b. \textbf{c} shows the score differences between two elements, similar to Fig. \ref{fig:high_low}c. Only significant differences are colored ($p<0.05$, corrected by FWER). \textbf{d,e} show element root-mean-square energy ratio or the predicted element emotion-induction effect of each audio in three datasets. Each line connects values of the two elements for an audio. If the proportion of the human-voice element is higher than that of the background-sound element, it is represented in red; otherwise, it is represented in blue. The prediction model is trained by the leave one out method. \textbf{f–h} Joint stepwise regression of voice and soundtrack features predicting emotion scores (behavioral: \textbf{f}; neural data in SEED: \textbf{g}; neural data in BAVE: \textbf{h}), using the same methodology as Fig. \ref{fig:high_low}d–f. Whiskers reflect variability (box-plot conventions), showing dataset/region-specific dominance of voice vs. soundtrack in emotion encoding.}
  \label{fig:bgm_vocal}
\end{figure*}

Behavioral data analysis reveals contrasting patterns: in the LIRIS dataset, background soundtracks outperform voices in both $H8$ and $H8_{PCA}$ (Fig. \ref{fig:bgm_vocal}b) and joint stepwise analysis (Fig. \ref{fig:bgm_vocal}f); While in SEED, the voice element yields higher emotion scores (Fig. \ref{fig:bgm_vocal}b, \ref{fig:bgm_vocal}f). Neural modeling of source-level activity of SEED (Fig. \ref{fig:bgm_vocal}c) shows voices dominate in regions like the language-related temporal cortex, with joint stepwise analyses highlighting voice primacy in cognition-related annotations and neural data of FP-rh (Fig. \ref{fig:bgm_vocal}f, \ref{fig:bgm_vocal}g). However, the limbic rACC-rh favors background sounds (Fig. \ref{fig:bgm_vocal}g), aligning with its role in unconscious emotion arousal from ambient acoustic contexts \cite{nummenmaa_emotional_2014}. In the BAVE dataset, voices elicit stronger emotion-related neural responses across most brain regions, particularly in the prefrontal cortex (Fig. \ref{fig:bgm_vocal}c, \ref{fig:bgm_vocal}h).

Dataset-specific dominance of voice or soundtrack elements prompts an investigation into stimulus-related factors. Computing root-mean-square energy of separated elements (Fig. \ref{fig:bgm_vocal}d) reveals LIRIS audios have higher background sound energy, SEED balances the energy of voice/sound, and BAVE minimizes background content.
To further validate the link between audio attributes and encoding effects on audio-level, we applied Leave-one-out (LOO) method to predict the element effect for each audio. Using human annotations for SEED/LIRIS and FP-rh neural data for BAVE, Fig. \ref{fig:bgm_vocal}e shows BAVE audios mainly trigger voice-mediated emotions, while LIRIS relies more on background sounds. The result of Leave-one-out effect predictions mirrors the trends of stimulus composition, linking element dominance to audio attribute heterogeneity.

\section{Discussion} \label{sec:discussion}
Unlike prior works in affective computing mainly focusing on static emotion recognition/prediction from physiological signals or stimulus features \cite{wang_video_2015,baveye_affective_2018,alarcao_emotions_2019,jiang_seedvii_2024,zheng_identifying_2019,zheng_investigating_2015}, our study offers novel insights into \textbf{the analysis of induction mechanisms} using emotional dynamics. Conventional EEG-based emotion recognition datasets provide millisecond-level EEG data paired with film-level emotion labels (e.g., positive, neutral, negative) \cite{zheng_investigating_2015,koelstra_deap_2012}. Many studies assigned the same emotional label to the entire EEG data corresponding to a film clip, segmenting it into shorter samples for model training \cite{zheng_identifying_2019,zhao_classification_2019,shen_contrastive_2022}. This approach ignores the dynamic changes in emotions during viewing, limiting the practical application of computational models. In contrast, our study employs behavioral markers and dynamic neural synchrony techniques to achieve multimodal dynamic emotion measurement, enabling cross-modal investigation with consistent results.
Moreover, while previous work \cite{zheng_identifying_2019,zhao_classification_2019,shen_contrastive_2022} has emphasized the importance of EEG signals from the temporal lobe in emotion recognition, this region's dual role in primary auditory function and high-level language processing, combined with the complex elements (e.g., dialogue, background music) in the audios, necessitates a more nuanced approach. Our study integrates audio stimulus representation and decomposition techniques to elucidate the emotion-elicitation mechanisms of various auditory elements, thereby enhancing our understanding of the brain's emotional encoding processes.

In addition, our work adds a complementary dimension to affective neuroscience by shifting the focus from \textbf{where} emotions are represented in the brain to \textbf{how} naturalistic auditory stimuli drive dynamic emotional responses. Prior neuroimaging studies \cite{koide-majima_distinct_2020,horikawa_neural_2020} have illuminated cortical emotion maps, characterizing spatial distributions of emotional states across the brain. However, these studies often lack mechanistic insights into how sensory features and stimulus components translate into neural and behavioral emotion dynamics.
Building on DNN encoding frameworks in vision and language \cite{li_dissecting_2023,caucheteux_evidence_2023,millet_realistic_2022,khosla_cortical_2021}, we introduce a hierarchical and parallel modeling approach to quantify how distinct stimulus components drive emotion-related responses, a dimension understudied in prior emotion neuroscience. Specifically, our work uniquely decomposes audio stimuli using Spleeter \cite{hennequin_spleeter_2020}, wav2vec 2.0 \cite{baevski_wav2vec_2020} and Hubert \cite{hsu_hubert_2021} to reveal: (1) auditory features from middle model layers outperform shallow acoustic cues in emotion encoding (Fig. \ref{fig:layerwise}, Supplementary Fig. 10), demonstrating prosodic and semantic abstraction as a key driver of emotional induction; (2) voices and soundtracks exhibit dataset-dependent emotion-evoking effects linked to stimulus energy distribution (Fig. \ref{fig:bgm_vocal}), bridging complex auditory scene analysis with emotion science. By integrating multi-scale feature mapping (handcrafted + DNN-derived), cross-dataset validation (SEED/LIRIS/BAVE), and neural-behavioral convergence (dynamic neural synchrony/behavioral annotations), our framework provides a stimulus-driven perspective on emotion arousal—complementing prior cortical emotion mapping studies by illuminating the transition from sensory input to neural/behavioral output in auditory-emotional processing.

Building on these scientific insights, the mechanistic framework developed here also holds practical implications for emotion-aware AI.
For emotion-aware AI, affective understanding—unlike static recognition—requires quantifying how cognitive elements combine to elicit emotions. Our framework, which decomposes audio into different stimuli components to model their differential emotional contributions, provides a foundation for fine-grained emotion analysis in industries like advertising. For example, such models could optimize audio design by disentangling the emotional impacts of vocal prosody vs. musical elements, aligning with the need for expert systems that inform content creation strategies.
However, translating these insights into real-world applications faces critical methodological challenges. 
First, the scarcity of time-aligned stimulus data and dynamic emotion annotation pairs hinders model training—an issue we address by expanding the annotation dataset at the behavioral level and using dynamic neural synchrony as a potential indicator of emotion arousal. Second, annotating the independent emotional impact of individual stimulus components (e.g., musical instruments in complex audio) remains prohibitively labor-intensive, necessitating computational approaches to infer such contributions indirectly. These challenges underscore the need for hybrid frameworks that integrate data-driven modeling with theoretical priors, while highlighting opportunities for cross-disciplinary innovation in emotion science.

Beyond these broader field challenges, the present work has specific limitations worthy of note. First, our analysis focuses on group-level averages, overlooking individual differences in auditory emotion encoding—an oversight highlighted by the dataset demographics: while BAVE/SEED primarily include Chinese participants and LIRIS features French speakers, cross-cultural generalizability remains untested. Second, although we decomposed audio into voices/soundtracks, we did not systematically control for stimulus categories (e.g., comedy vs. tragedy in SEED/BAVE, six genres in LIRIS). Whether emotion encoding mechanisms vary across film types—such as the distinct roles of vocal humor in comedies vs. orchestral cues in tragedies—requires dedicated investigation. Additionally, the spatial resolution of EEG source localization (1 cm) may limit precise neural regionalization, though our use of both raw EEG and source-localized activity (68 regions) mitigates this limitation. Future research could incorporate diverse cultural cohorts, model individual differences, and design genre-controlled experiments to further investigate the auditory emotion encoding process.

\section{Conclusion}
This study integrates affective neuroscience and computing to unveil hierarchical mechanisms of auditory-emotion encoding, constructing a neurobiologically informed computational framework that maps naturalistic auditory inputs to dynamic behavioral/neural responses across SEED, LIRIS, and BAVE datasets. Guided by neurobiological principles, we decompose audio into multilevel features (via classical feature extraction algorithms or wav2vec 2.0/Hubert) from the original or the isolated human voice/background soundtrack elements, yielding three key contributions. First, semantic representations dominate emotion encoding, outperforming acoustic features with stronger mappings to behavioral annotations and dynamic neural synchrony across most brain regions ($p < 0.05$), particularly in right hemisphere areas like the rostral anterior cingulate cortex and the frontal pole cortex. Second, wav2vec 2.0/Hubert’s middle layers (7–14) balance acoustic-semantic information to surpass final layers in emotion induction robustness. Third, voices and soundtracks exhibit dataset-dependent biases aligned with energy distribution (e.g., LIRIS favors soundtracks), with voices dominating prefrontal/temporal activity and soundtracks excelling in limbic regions.

These findings not only elucidate how the brain integrates auditory hierarchy and stimulus ecology to drive emotion arousal dynamics but also advance emotion-aware AI. As highlighted by Picard’s vision for affective computing \cite{picard_affective_2000}, this work provides a neurocomputational foundation for adaptive systems, bridging artificial intelligence with human emotion science to enable more sophisticated modeling of audio-affective interactions.

\bibliographystyle{IEEEtran}
\bibliography{Zotero}

\begin{IEEEbiography}[{\includegraphics[width=1in,height=1.25in,clip,keepaspectratio]{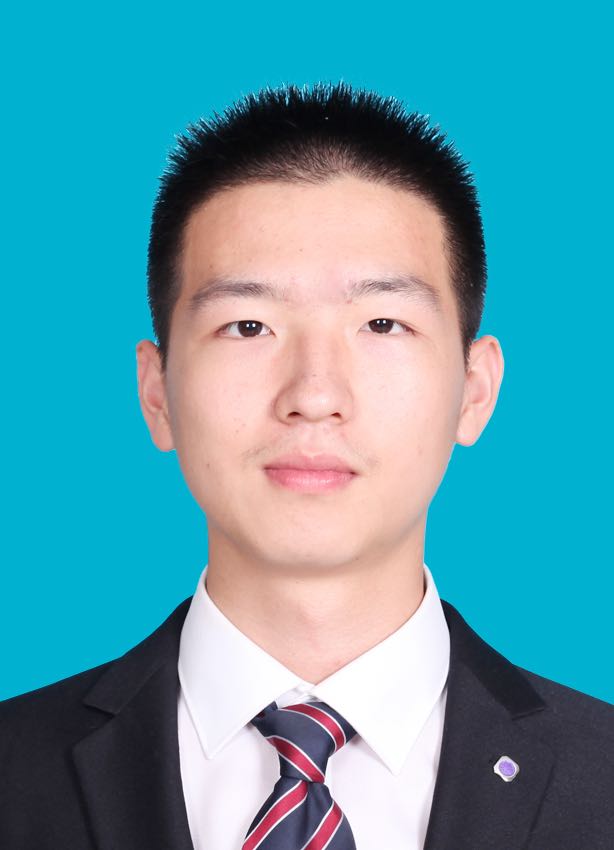}}]{Guandong Pan} obtained his Bachelor of Engineering degree from the School of Computer Science and Engineering at Beihang University from 2016 to 2020. Since 2020, he has been pursuing a doctoral degree at Beihang University. His current research interests include dynamic emotion decoding, and affective understanding via neural encoding.
\end{IEEEbiography}

\begin{IEEEbiography}[{\includegraphics[width=1in,height=1.25in,clip,keepaspectratio]{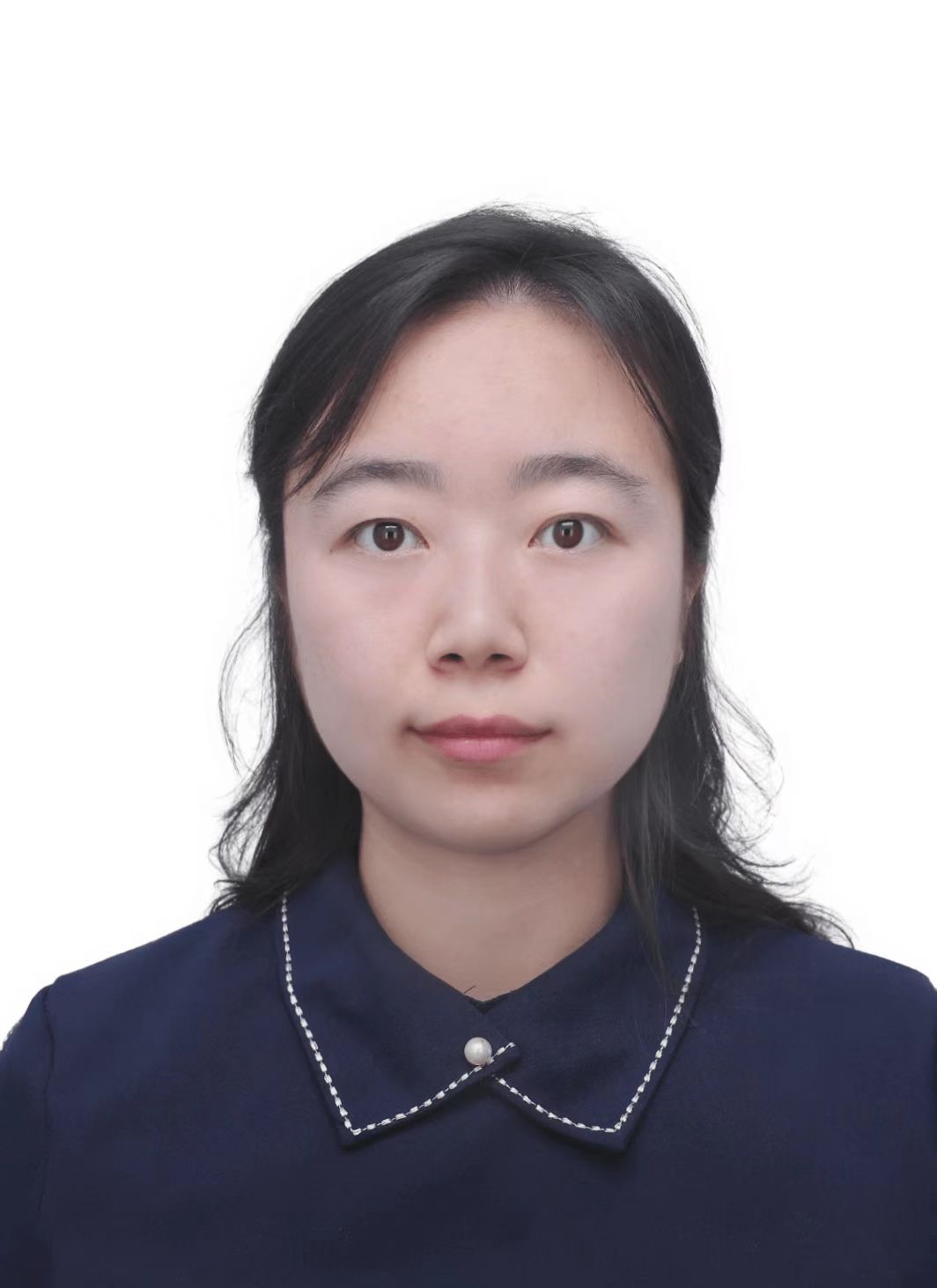}}]{Yaqian Yang} earned a dual Bachelor of Science degree and Bachelor of Financial Engineering degree from the University of Science and Technology Beijing between 2012 and 2016. From 2016 to 2019, she obtained a Master of Science degree from the School of Mathematical Sciences at Beihang University. She received her Doctor of Science from Beihang University in 2024. Her research findings have been published in journals such as Nature Communications and Communications Biology. Currently, she serves as a postdoctoral fellow at the Institute of Artificial Intelligence, Beihang University. Her current research continues to explore the complex topological structure and dynamic processes of brain networks, multimodal cross-scale coupling mechanisms, neural bases of individual specificity and cognitive functions, and mathematical modeling of complex spatiotemporal brain patterns. 
\end{IEEEbiography}

\begin{IEEEbiography}[{\includegraphics[width=1in,height=1.25in,clip,keepaspectratio]{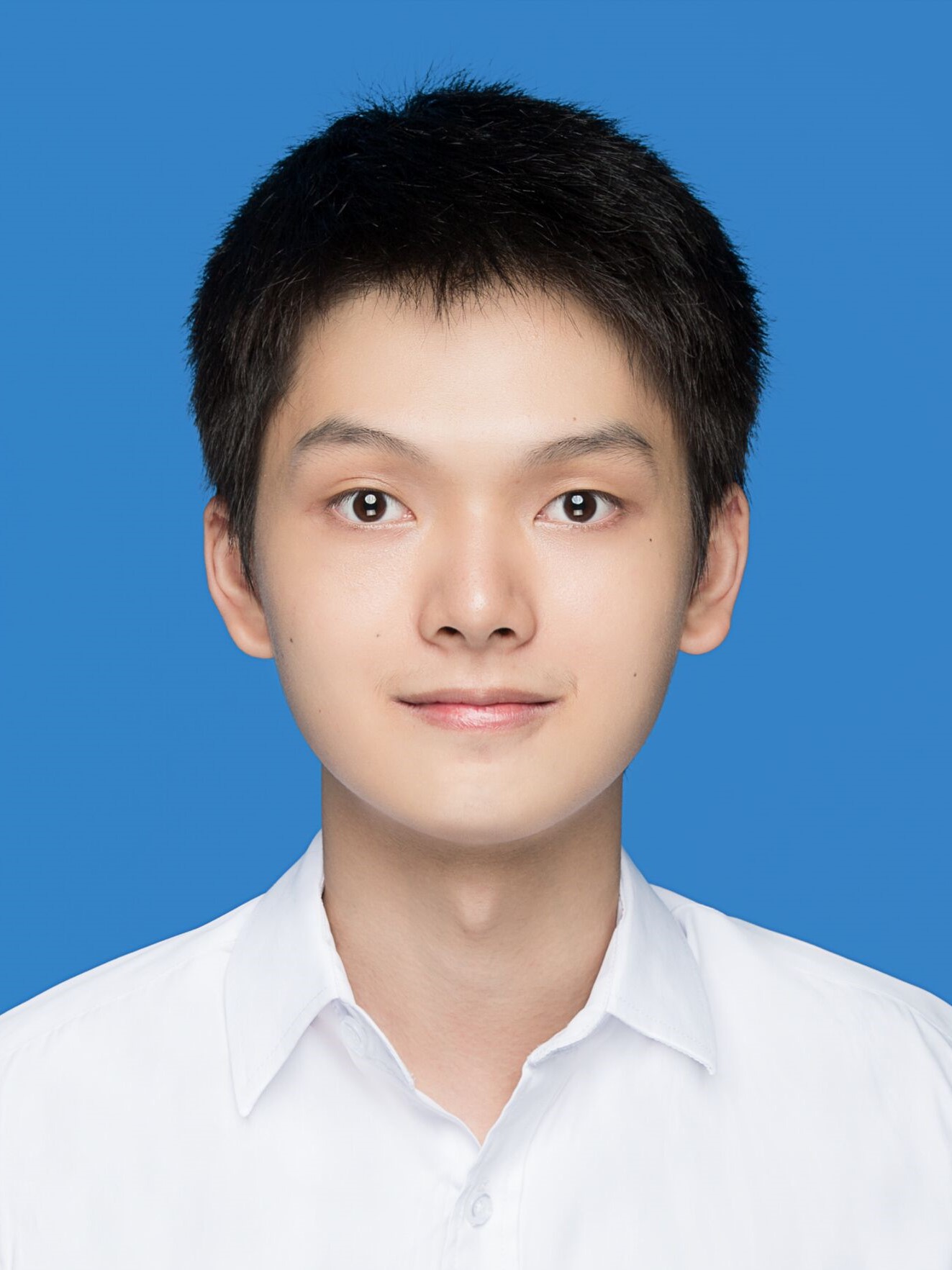}}]{Shi Chen} received BS degree from Beihang University, China in 2024. He is currently a doctoral student at the Institute of Artificial Intelligence, Beihang University. His research interests include biomedical signal processing, brain computer interface, and machine learning.
\end{IEEEbiography}

\begin{IEEEbiography}[{\includegraphics[width=1in,height=1.25in,clip,keepaspectratio]{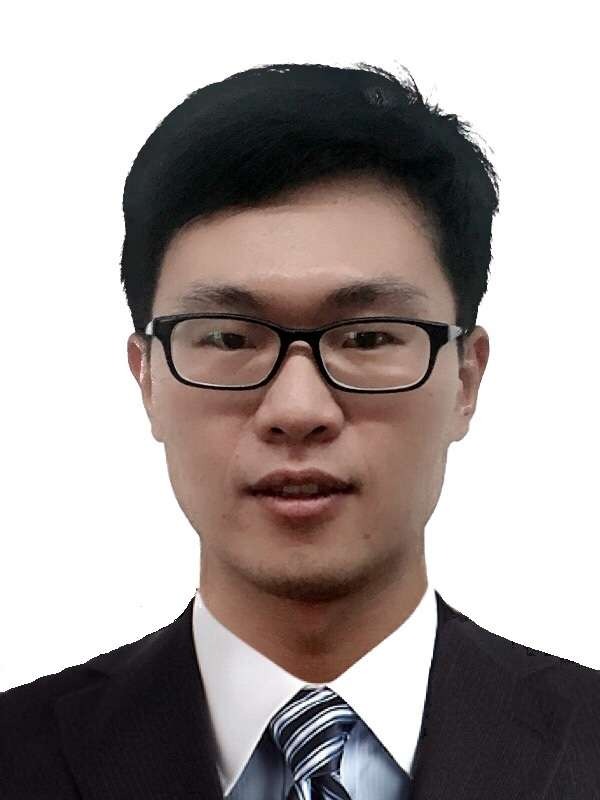}}]{Xin Wang} received BS degree from Beihang University, China in 2015. After, he received his Ph.D. degree from Beihang University, China in 2021. Now, he is an Associate Professor at the Institute of Artificial Intelligence, Beihang University. He has published over 20 SCI papers in top journals like Physical Review X and Nature Communications, receiving honors such as PRX Highlights Paper and Chaos Editor's Pick. His research interests include big data, complex social systems, swarm intelligence, and brain-inspired AI. He is a guest editor of Entropy.
\end{IEEEbiography}

\begin{IEEEbiography}[{\includegraphics[width=1in,height=1.25in,clip,keepaspectratio]{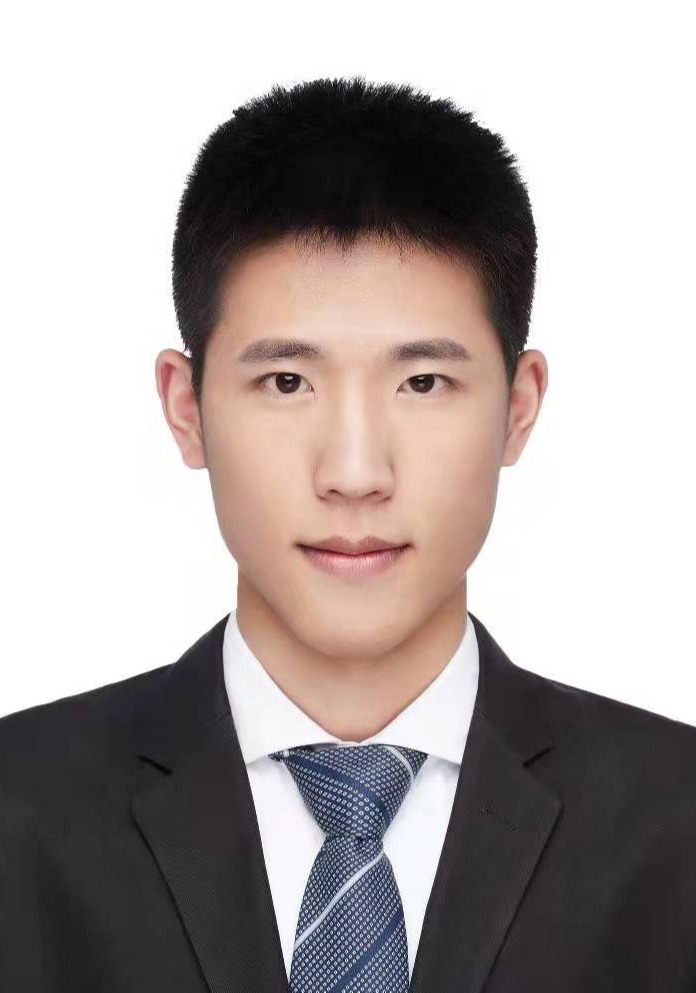}}]{Longzhao Liu} received a bachelor's degree and then a doctoral degree from the School of Mathematical Sciences, Beihang University. He had an exchange at the Center for Complex Systems Research, Northwestern University, US, from 2019 to 2020. Currently, he works as an assistant professor and master's supervisor at the Institute of Artificial Intelligence, Beihang University. He has published 6 papers in journals like New J. Phys. and J. Stat. Mech.-Theory Exp. His research interests focus on complex systems, network science, group behavior and idea evolution, swarm intelligence, and dissemination dynamics, exploring microscopic dynamics to explain macroscopic phenomena in complex systems.
\end{IEEEbiography}

\begin{IEEEbiography}[{\includegraphics[width=1in,height=1.25in,clip,keepaspectratio]{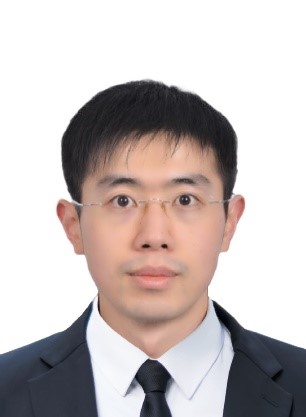}}]{Hongwei Zheng} earned his Doctor degree from the Ohio State University, department of economics. He is a senior researcher at the Beijing Academy of Block-Chain and Edge Computing. He has built many IT platforms using block-chain and distributed ledger technology with an emphasis on serving carbon neutrality and privacy preserving. He’s research interest includes private computing, complex networks and big data. He is a member of IEEE and CIC.
\end{IEEEbiography}

\begin{IEEEbiography}[{\includegraphics[width=1in,height=1.25in,clip,keepaspectratio]{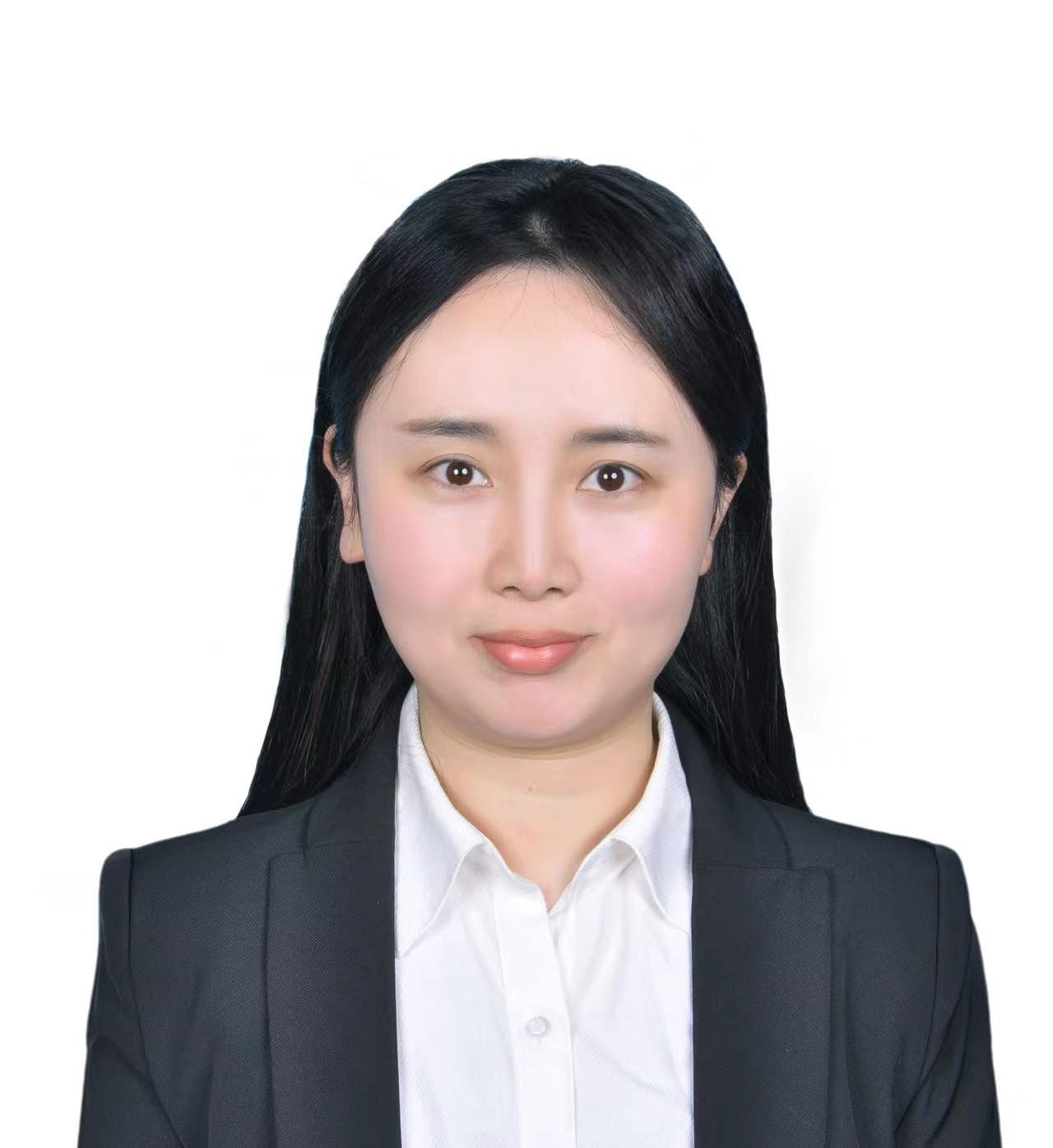}}]{Shaoting Tang} earned a Bachelor of Science degree in Information and Computational Science from the Department of Applied Mathematics at Dalian University of Technology from 2002 to 2006. Subsequently, from 2006 to 2011, she obtained a Doctorate of Science from the School of Mathematics and Systems Science at Beihang University. Currently, she serves as the deputy director of both the State Key Laboratory of Software Development Environment and the Key Laboratory of Mathematics, Informatics, and Behavior, Ministry of Education.  Her current research interests lie in artificial intelligence, complex systems, brain cognition, and brain networks. 
\end{IEEEbiography}


\vfill

\end{document}